# Electronic transport property of PbS nanowire devices

Jiarui Yin[1*], Ming Zheng[2], Kuo Hai[2]
1.Department of Chemical Engineering, School of Chemical Engineering and Pharmacy,
Wuhan Institute of Technology, Wuhan 430205,China
2.Department of Physics and Key Laboratory of Low-dimensional Quantum Structures and Quantum Control of Ministry of Education, and Synergetic Innovation Center for Quantum Effects and Applications,
Hunan Normal University, Changsha 410081, China

Lead sulfide is an important photosensitive material, and its photoelectric properties have received widespread attention. We completed the preparation of PbS nanowires and single PbS nanowire devices, then we conducted electrical performance tests on the PbS nanowire devices. We found that a single lead sulfide nanowire device exhibits memristive properties that depend on the applied voltage and the power density of light. From this we studied the electrical transport properties of single lead sulfide nanodevices.

## Introduction

Lead sulfide (PbS) is a direct band gap material with cubic crystal phase. Lead sulfide have a narrow band gap and larger exciton Bohr radius, which makes lead sulfide with photoconductivity and photovoltaic phenomena[1-4]. In recent years, semiconductor nanowires (NWs) have widely investigated for their potential deployment in the electronics and optoelectronics technologies[5]. Their large aspect ratio makes them an interesting tool with which to investigate the nanoscale properties of materials. PbS attracts a lot of attention, because of the narrow gap (0.41eV) and its larger Bohr radius (18nm). A range of promising devices including field effect transistors[6], photodetectors[7,8] and solar cells[9,10] have been demonstrated. There has been a surge in nanowire research because Vapour-Liquid-Solid synthesis allowed for the largescale synthesis of nanowires with diameters beyond the limits of conventional lithography[11,12]. However, the method requires high growth temperatures accompanied by a precise control of the pressure or flow of the reaction gases[13,14]. Therefore, the research focus has shifted towards cheaper and more facile approaches involving the growth of nanowires from a liquid phase. In previous papers, there are a few reports of liquid phase growth of PbS nanowires[15,16]. In this work, the PbS nanowire based on the Pb nanowires were prepared successfully. We also fabricated devices using single PbS nanowire. We studied the effect of light on the current and memristive characteristics of the device and discussed the electrical transport properties of Pbs devices.

*Email: ron.khai@gmail.com;

## Experimental details

The PbS nanowires based on the Pb nanowires were fabricated by solvothermal method[17]. The thiourea and Pb nanowires were dissolved in 50mL ethylene glycol respectively. Ethylene glycol（100mL） take place in a flask and heated to 180℃. Then, the thiourea solution and the solution with Pb nanowires were injected into the hot ethylene glycol solution slowly. The reaction was maintained at 160℃ for 1.5h. After the completion of the reaction, the reaction mixture was quickly cooled to room temperature by means of a water bath. The PbS nanowires were precipitated from the solution by centrifuged, dried. After that photolithography was used to deposit a combination of 10/90 nm of Ti / Au electrodes onto a highly p-doped silicon substrate, covered by a thermally grown layer of 300nm SiO2. All the devices were stored in ambient and room temperature. Without additional processing, the contact between the Au electrode and the PbS nanowire is an ohmic contact.

## Results and discussion

The morphology of Pb nanowires and PbS nanowires was characterized by scanning electron microscope (SEM). The SEM image of Pb nanowires and PbS nanowires are presented in Figure 1(a),1(b). The results show that the Pb nanowires and PbS nanowires have smooth surfaces. The diameters of Pb nanowires and PbS nanowires are 100~300nm and 200~500nm, respectively. The X-ray diffraction (XRD) pattern of the Pb nanowires and PbS nanowires are shown in Figure 1(c). It is found that the peaks closely match the XRD pattern obtained for PDF 04-0686 and PDF 01-0880, indicating that the as-grown Pb nanowires and PbS nanowires have high degree of crystallinity. Figure 1(d) is the energy dispersive X-ray spectroscopy (EDS) pattern of PbS nanowires. The inset is the atomic ratio of each element in the sample. The result shows that the lead-sulfur element ratio was close to 1:1, further indicating that the sample we grew was PbS.

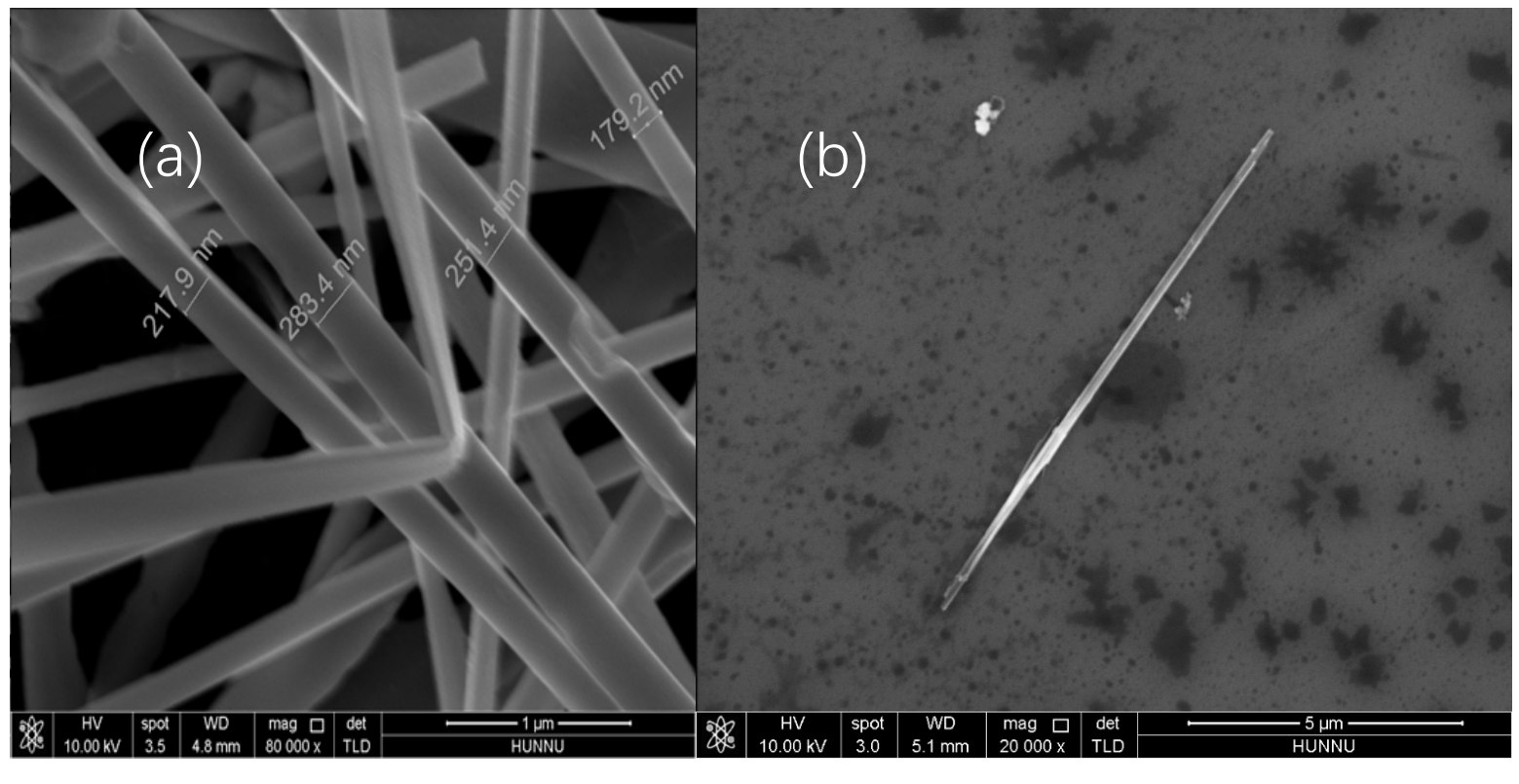

Figure 1. (a) SEM image of Pb NWs. (b) SEM image of PbS NWs.

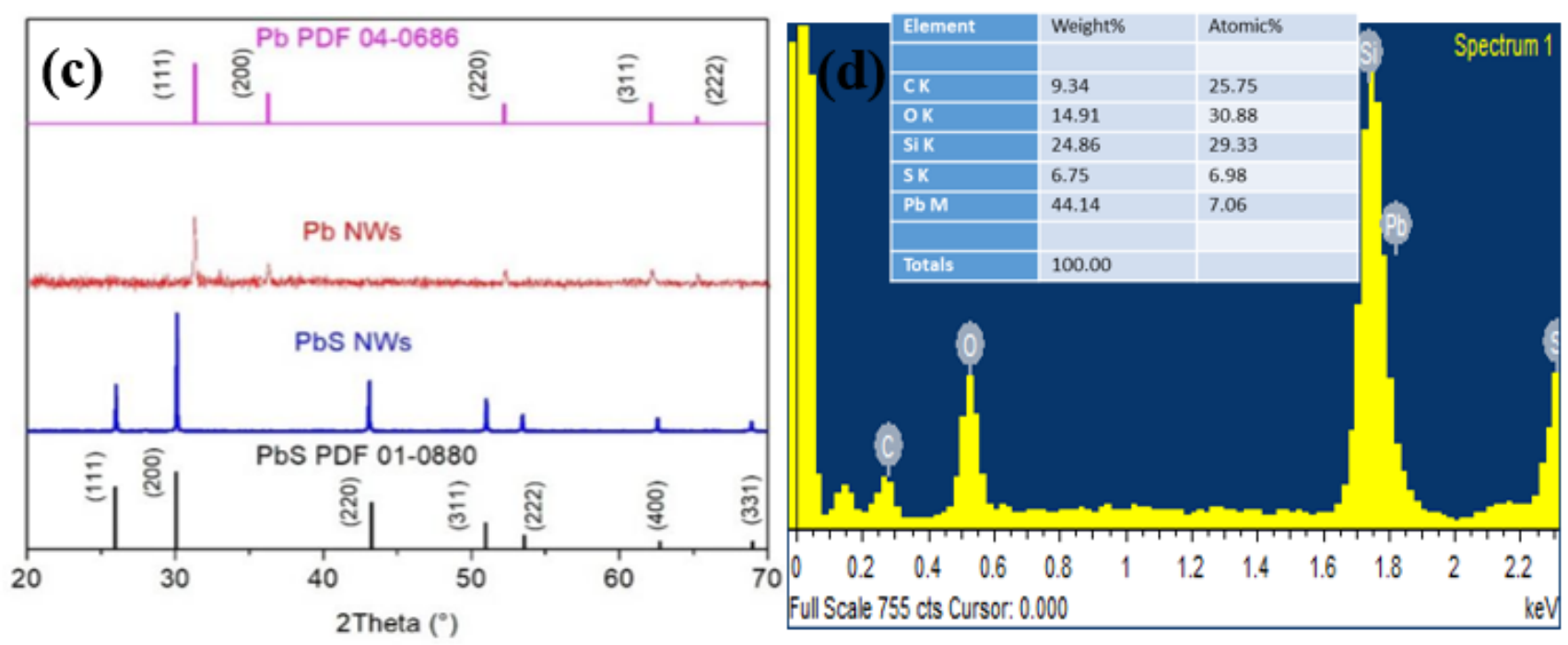

Figure 1. (c) XRD pattern of Pb NWs and PbS NWs. (d) EDS pattern of PbS NWs.

The PbS nanowires were spin-coated onto a silicon substrate. The Ti/Au electrodes had been patterned using standard photolithography techniques. The SEM image of a PbS nanowires device is shown in Figure 2(a). We tested the I-V curve of the device under low voltage, as shown in Figure 2(b), the curve is almost linear. It reveals that there is an ohmic contact between the PbS nanowire and electrodes. Figure 2(c), 2(d) are the I-V curve of devices with different voltage. We found that the I-V curves completely coincide under the voltage of -1~1V. As the applied voltage increases, the I-V curve gradually appears hysteresis, which shows that PbS nanowires have memristive properties. In our view, the cause of memristive mechanism is that there are a lot of defects in the PbS nanowires. When an external electric field is applied, the injected electrons will be trapped by the defects. As the voltage increases, the defects are gradually completely occupied by electrons, and the electron concentration in the nanowire will rise sharply.

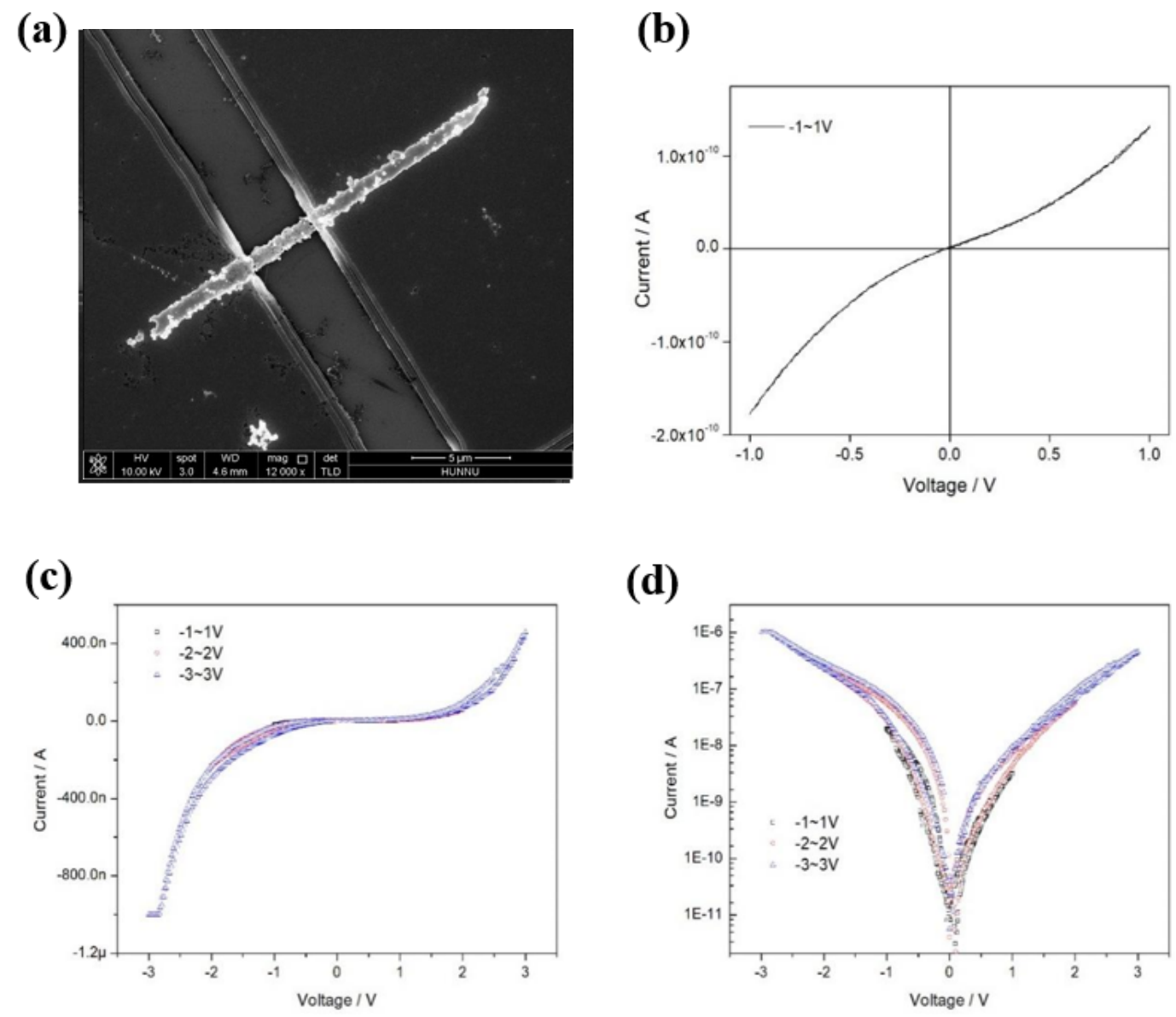

Figure 2. (a) SEM image of PbS NWs device. (b) I-V curve of device for low voltage. (c) I-V curve of device with different scanning voltage. (d) I-V curve of device in the dark with different scanning voltage plotted on a log-scale.

Compared with bulk materials, one-dimensional nanomaterials have a stable carrier transport channel with large aspect ratio and large specific surface area, which can achieve high-efficiency carrier transport. PbS nanomaterials have shown particular promise for optoelectronics and electronics applications due to their extensive band gap tunability and earth-abundant availability. The stability of device performance is shown in Figure 3(a). With the increase of the scanning numbers, the hysteresis of the I-V curve becomes more and more obvious. It reveals that the memristive performance of the device has increased. We compared the electrical performance of PbS NWs devices under dark and normal light in Figure 3(b),3(c). When the scanning voltage is -2~2V, the current of the device in the normal light is obviously smaller than that in the dark. In our view, the effect of external light will enhance the desorption of water and oxygen molecules. Desorption results in reduced electron concentration of the NW. When switching from dark to normal light, the hysteresis of the I-V curve of the device was disappeared. The defects and the desorption of water and oxygen molecules make the electron concentration in the nanowires lower[18,19]. To make sharply changes in current, you need to apply a higher voltage to the NW. Therefore, the memristive performance of the devices decreases with the increase of the power density of light. The electrical performance of devices with different scanning voltages in normal light is shown in Figure 3(d). As the scanning voltage increases, the hysteresis of the I-V curve becomes larger. It indicated that the memristive performance increases with the increase of scanning voltage.

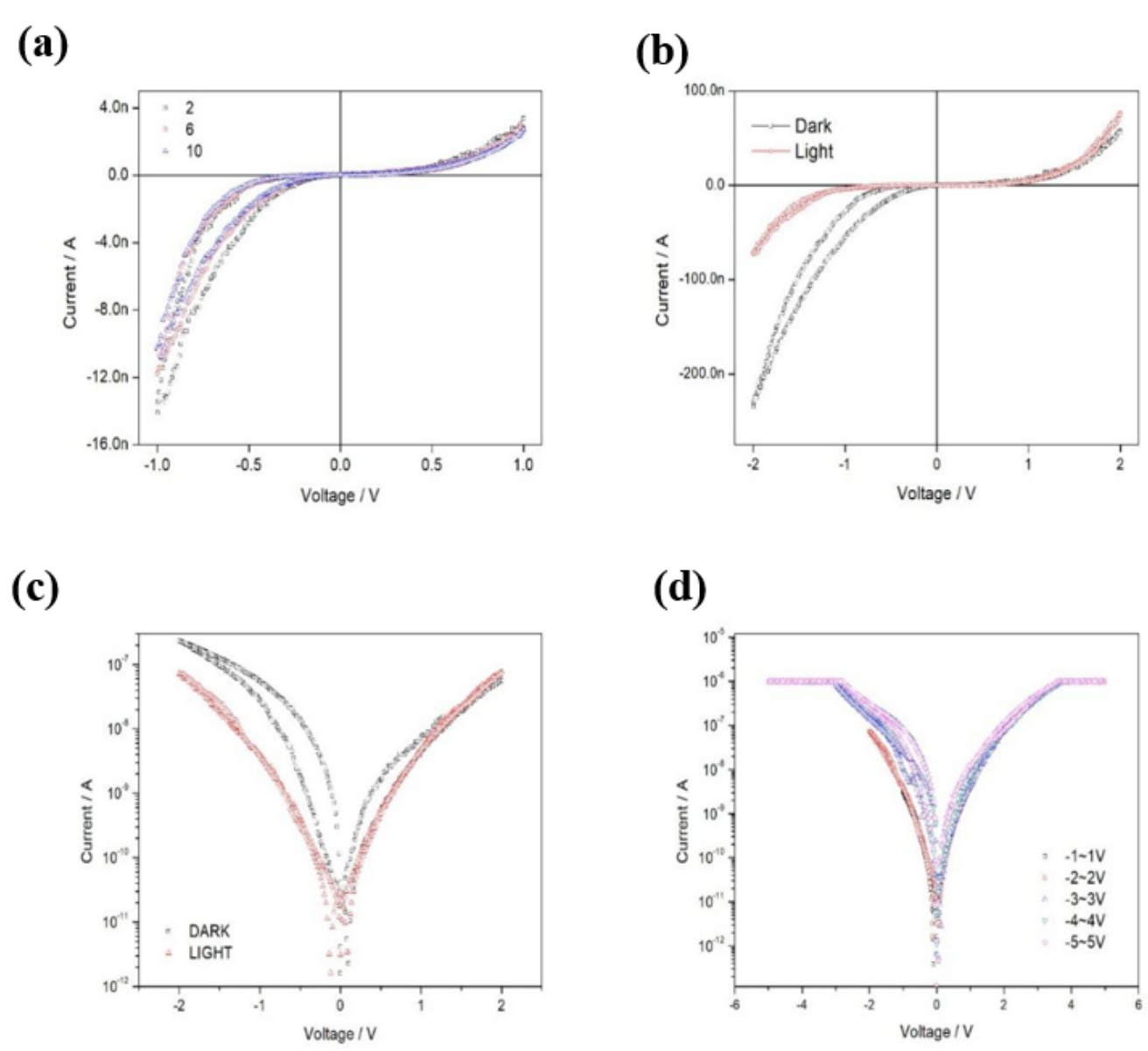

Figure 3. (a) I-V curve of device for scanning numbers. (b) I-V curve of device in the dark or in normal light for -2~2V. (c) I-V curve of device in the dark or in normal light for -2~2V plotted on a log-scale. (d) I-V curve of device in normal light with different scanning voltage plotted on a log-scale.

We scanned the PbS nanowire device with a low voltage, and its I-V curves almost completely overlapped, indicating that the PbS nanowire device's own resistance did not change during the scanning process and had no memristive performance. As the scanning voltage increases, the I-V curves of the PbS nanowire device do not overlap in the forward and reverse scans, indicating that the conductivity of the PbS nanowire changes during the scanning process, showing memristive properties. The higher the scanning voltage applied, the more obvious the hysteresis I-V curve of the PbS nanowire device shows, and its memristive performance get stronger. At the same voltage, the number of scans also affects the electrical properties of PbS nanowire devices. By comparing the I-V curves of PbS nanowire devices under dark field and bright field, we can see that the current value in bright field is significantly smaller than that in dark field under the same voltage. In the bright field, the electrons in the PbS nanowires absorb the energy of light to form photogenerated free electrons and participate in conduction. In the bright field, the concentration of electrons in PbS nanowires increases, but the current value decreases instead, indicating that the conductivity type of PbS nanowires is P-type, and the majority of carriers are holes. The photogenerated free electrons will recombine with the holes in the nanowire itself, which will reduce the number of carriers participating in the conduction in the PbS nanowire and affect its electrical properties.

## Conclusion

In summary, solution-processed PbS nanowires have been successfully synthesized·. XRD and SEM analysis indicated that PbS nanowires have a high degree of crystallinity. The devices exhibited memristive property. The memristive performance increases with the increase of the scanning voltage, and decreases with the increase of the power density of light. It was found that the current of the device became smaller in the bright field, indicating that the lead sulfide nanowire device was hole-conducting.